\documentclass[conference,10pt]{IEEEtran}
\IEEEoverridecommandlockouts
\usepackage{cite}
\usepackage{amsmath,amssymb,amsfonts}
\usepackage[nolist]{acronym}
\usepackage{hyperref}
\usepackage{scalerel}
\usepackage{graphicx}
\usepackage{nicefrac}
\usepackage{xcolor}
\usepackage{circuitikz}

\usetikzlibrary{calc}
\usepgflibrary{shadings}
\usepackage{pgfplots}
\usepgfplotslibrary{groupplots}

\pgfplotsset{compat=1.15}

\makeatletter
\let\MYcaption\@makecaption
\makeatother

\usepackage[font=footnotesize]{subcaption}

\makeatletter
\let\@makecaption\MYcaption
\makeatother

\DeclareMathOperator*{\argmax}{arg\,max}

\usetikzlibrary{positioning}
\usetikzlibrary{svg.path}

\def\BibTeX{{\rm B\kern-.05em{\sc i\kern-.025em b}\kern-.08em
    T\kern-.1667em\lower.7ex\hbox{E}\kern-.125emX}}

\definecolor{orcidlogocol}{HTML}{A6CE39}
\tikzset{
  orcidlogo/.pic={
    \fill[orcidlogocol] svg{M256,128c0,70.7-57.3,128-128,128C57.3,256,0,198.7,0,128C0,57.3,57.3,0,128,0C198.7,0,256,57.3,256,128z};
    \fill[white] svg{M86.3,186.2H70.9V79.1h15.4v48.4V186.2z}
                 svg{M108.9,79.1h41.6c39.6,0,57,28.3,57,53.6c0,27.5-21.5,53.6-56.8,53.6h-41.8V79.1z M124.3,172.4h24.5c34.9,0,42.9-26.5,42.9-39.7c0-21.5-13.7-39.7-43.7-39.7h-23.7V172.4z}
                 svg{M88.7,56.8c0,5.5-4.5,10.1-10.1,10.1c-5.6,0-10.1-4.6-10.1-10.1c0-5.6,4.5-10.1,10.1-10.1C84.2,46.7,88.7,51.3,88.7,56.8z};
  }
}

\newcommand\orcidicon[1]{\href{https://orcid.org/#1}{\mbox{\scalerel*{
\begin{tikzpicture}[yscale=-1,transform shape]
\pic{orcidlogo};
\end{tikzpicture}
}{|}}}}

\begin{acronym}
 \acro{R2M}{raw 2\textsuperscript{nd} moment}
 \acro{CSI}{channel state information}
 \acro{UE}{user equipment}
 \acro{EM}{electromagnetic}
 \acro{UL}{uplink}
 \acro{BS}{base station}
 \acro{TDD}{time division duplex}
 \acro{FDD}{frequency division duplex}
 \acro{ECC}{error-correcting code}
 \acro{MLD}{maximum likelihood decoding}
 \acro{HDD}{hard decision decoding}
 \acro{IF}{intermediate frequency}
 \acro{RF}{radio frequency}
 \acro{SDD}{soft decision decoding}
 \acro{NND}{neural network decoding}
 \acro{CNN}{convolutional neural network}
 \acro{ML}{maximum likelihood}
 \acro{GPU}{graphical processing unit}
 \acro{BP}{belief propagation}
 \acro{LTE}{Long Term Evolution}
 \acro{BER}{bit error rate}
 \acro{SNR}{signal-to-noise-ratio}
 \acro{ReLU}{rectified linear unit}
 \acro{BPSK}{binary phase shift keying}
 \acro{QPSK}{quadrature phase shift keying}
 \acro{AWGN}{additive white Gaussian noise}
 \acro{MSE}{mean squared error}
 \acro{LLR}{log-likelihood ratio}
 \acro{MAP}{maximum a posteriori}
 \acro{NVE}{normalized validation error}
 \acro{BCE}{binary cross-entropy}
 \acro{CE}{cross-entropy}
 \acro{BLER}{block error rate}
 \acro{SQR}{signal-to-quantisation-noise-ratio}
 \acro{MIMO}{multiple-input multiple-output}
 \acro{mMIMO}{massive multiple-input multiple-output}
 \acro{OFDM}{orthogonal frequency division multiplex}
 \acro{RF}{radio frequency}
 \acro{LoS}{line of sight}
 \acro{NLoS}{non-line of sight}
 \acro{NMSE}{normalized mean squared error}
 \acro{CFO}{carrier frequency offset}
 \acro{SFO}{sampling frequency offset}
 \acro{IPS}{indoor positioning system}
 \acro{TRIPS}{time-reversal IPS}
 \acro{RSSI}{received signal strength indicator}
 \acro{MIMO}{multiple-input multiple-output}
 \acro{ENoB}{effective number of bits}
 \acro{AGC}{automatic gain control}
 \acro{ADC}{analog to digital converter}
 \acro{ADCs}{analog to digital converters}
 \acro{FB}{front bandpass}
 \acro{FPGA}{field programmable gate array}
 \acro{JSDM}{Joint Spatial Division and Multiplexing}
 \acro{NN}{neural network}
 \acro{IF}{intermediate frequency}
 \acro{LoS}{line-of-sight}
 \acro{NLoS}{non-line-of-sight}
 \acro{DSP}{digital signal processing}
 \acro{AFE}{analog front end}
 \acro{SQNR}{signal-to-quantisation-noise-ratio}
 \acro{SINR}{signal-to-interference-noise-ratio}
 \acro{ENoB}{effective number of bits}
 \acro{PCB}{printed circuit board}
 \acro{EVM}{error vector mangnitude}
 \acro{CDF}{cumulative distribution function}
 \acro{MRC}{maximum ratio combining}
 \acro{MRP}{maximum ratio precoding}
 \acro{MRT}{maximum ratio transmission}
 \acro{DeepL}{deep-learning}
 \acro{DL}{downlink}
 \acro{SISO}{single-input single-output}
 \acro{SGD}{stochastic gradient descent}
 \acro{CP}{cyclic prefix}
 \acro{MISO}{Multiple Input Single Output}
 \acro{LMMSE}{linear minimum mean square error}
 \acro{ZF}{zero forcing}
 \acro{USRP}{universal software radio peripheral}
 \acro{RNN}{recurrent neural network}
 \acro{GRU}{gated recurrent unit}
 \acro{LSTM}{long short-term memory}
 \acro{NTM}{neural turing machine}
 \acro{DNC}{differentiable neural computer}
 \acro{TCN}{temporal convolutional network}
 \acro{FCL}{fully connected layer}
 \acro{MANN}{memory augmented neural network}
 \acro{RNN}{recurrent neural network}
 \acro{DNN}{deep neural network}
 \acro{FIR}{finite impulse response}
 \acro{BPTT}{back-propagation through time}
 \acro{GAN}{generative adversarial network}
 \acro{ELU}{exponential linear unit}
 \acro{tanh}{hyperbolic tangent}
 \acro{BICM}{bit-interleaved coded modulation}
 \acro{OTA}{over-the-air}
 \acro{IM}{intensity modulation}
 \acro{DD}{direct detection}
 \acro{RL}{reinforcement learning}
 \acro{SDR}{software-defined radio}
 \acro{WGAN}{Wasserstein generative adversarial network}
 \acro{BMD}{bit-metric decoding}
 \acro{BMI}{bit-wise mutual information}
 \acro{LDPC}{low-density parity-check}
 \acro{IDD}{iterative demapping and decoding}
 \acro{JSD}{Jensen-Shannon divergence}
 \acro{MMSE}{minimum mean square error}
 \acro{FFT}{fast Fourier transform}
 \acro{DFT}{discrete Fourier transform}
 \acro{IFFT}{inverse fast Fourier transform}
 \acro{QAM}{quadrature amplitude modulation}
 \acro{EMD}{earth mover's distance}
 \acro{TDL}{tapped delay line}
 \acro{KL}{Kullback-Leibler}
 \acro{PRACH}{physical random access channel}
 \acro{URLLC}{ultra-reliable low-latency communication}
 \acro{ANOMA}{asynchronous non-orthogonal multiple access}
 \acro{FEC}{forward error correction}
 \acro{PAPR}{peak-to-average power ratio}
 \acro{APP}{a posteriori probability}
 \acro{COTS}{commercial off-the-shelf}
 \acro{PLL}{phase locked loop}
 \acro{STO}{sampling time offset}
 \acro{SFO}{sampling frequency offset}
 \acro{CFO}{carrier frequency offset}
 \acro{CPO}{carrier phase offset}
 \acro{CSI}{channel state information}
 \acro{GNSS}{global navigation satellite system}
 \acro{ELAA}{extremely large aperture array}
 \acro{UE}{user equipment}
 \acro{DICHASUS}{\underline{Di}stributed \underline{Cha}nnel \underline{S}ounder by \underline{U}niversity of \underline{S}tuttgart}
 \acro{JCaS}{Joint Communication and Sensing}
 \acro{CT}{continuity}
 \acro{TW}{trustworthiness}
 \acro{KS}{Kruskal's stress}
 \acro{PCA}{principal component analysis}
 \acro{AoA}{angle of arrival}
 \acro{ToA}{time of arrival}
 \acro{TDoA}{time difference of arrival}
 \acro{ToT}{time of transmission}
 \acro{RD}{Rajski's distance}
 \acro{ADP}{angle-delay profile}
 \acro{MPC}{multipath component}
 \acro{CIR}{channel impulse response}
 \acro{MAE}{mean absolute error}
 \acro{MDS}{multidimensional scaling}
 \acro{t-SNE}{t-distributed stochastic neighbor embedding}
 \acro{SM}{Sammon's mapping}
 \acro{CIRA}{channel impulse response amplitude}
 \acro{CS}{cosine similarity}
 \acro{FCF}{forward charting function}
 \acro{CMD}{correlation matrix distance}
 \acro{UWB}{ultra-wideband}
 \acro{MUSIC}{MUltiple SIgnal Classification}
 \acro{FBCM}{forward-backward correlation matrix}
 \acro{MDL}{minimum descriptive length}
 \acro{BFGS}{Broyden-Fletcher-Goldfarb-Shanno}
 \acro{UPA}{uniform planar array}
 \acro{RMS}{root mean square}
 \acro{DRMS}{distance root mean square}
 \acro{CEP}{circular error probable}
 \acro{R95}{95\textsuperscript{th} percentile radius}
 \acro{eCDF}{empirical cumulative distribution function}
 \acro{L-LTF}{legacy long training field}
 \acro{HT-LTF}{high throughput long training field}
 \acro{DOI}{Digital Object Identifier}
 \acro{PA}{Power Amplifier}
 \acro{BSSID}{Basic Service Set Identification}
 \acro{RAN}{Radio Access Network}
 \acro{PCC}{passive channel charting}
 \acro{ISAC}{integrated sensing and communication}
 \acro{CRAP}{Clutter Removal with Acquisitions Under Phase Noise}
 \acro{RCS}{radar cross section}
\end{acronym}

\definecolor{mittelblau}{RGB}{0, 126, 198}
\definecolor{violettblau}{cmyk}{0.9, 0.6, 0, 0}
\definecolor{rot}{RGB}{238, 28 35}
\definecolor{apfelgruen}{RGB}{140, 198, 62}
\definecolor{gelb}{RGB}{1, 221, 0}
\definecolor{orange}{RGB}{244, 111, 33}
\definecolor{pink}{RGB}{237, 0, 140}
\definecolor{lila}{RGB}{128, 10, 145}
\definecolor{hellgrau}{RGB}{224, 224, 224}
\definecolor{mittelgrau}{RGB}{128, 128, 128}
\definecolor{dunkelgrau}{RGB}{80,80,80}
\definecolor{anthrazit}{RGB}{19, 31, 31}

\begin{document}

\title{Passive Channel Charting: Locating Passive Targets using Wi-Fi Channel State Information
\thanks{This work is supported by the German Federal Ministry of Education and Research (BMBF) within the projects Open6GHub (grant no. 16KISK019) and KOMSENS-6G (grant no. 16KISK113).}}

\author{\IEEEauthorblockN{Florian Euchner\textsuperscript{\orcidicon{0000-0002-8090-1188}}, David Kellner\textsuperscript{\orcidicon{0009-0009-5054-582X}}, Phillip Stephan\textsuperscript{\orcidicon{0009-0007-4036-668X}}, Stephan ten Brink\textsuperscript{\orcidicon{0000-0003-1502-2571}} \\}
\IEEEauthorblockA{
Institute of Telecommunications, Pfaffenwaldring 47, University of  Stuttgart, 70569 Stuttgart, Germany \\ \{euchner,stephan,tenbrink\}@inue.uni-stuttgart.de
}
}

\newpage

\maketitle

\begin{abstract}
We propose passive channel charting, an extension of channel charting to passive target localization.
As in conventional channel charting, we follow a dimensionality reduction approach to reconstruct a physically interpretable map of target positions from similarities in high-dimensional channel state information.
We show that algorithms and neural network architectures developed in the context of channel charting with active mobile transmitters can be straightforwardly applied to the passive case, where we assume a scenario with static transmitters and receivers and a mobile target.
We evaluate our method on a channel state information dataset collected indoors with a distributed setup of ESPARGOS Wi-Fi sensing antenna arrays.
This scenario can be interpreted as either a multi-static or passive radar system.
We demonstrate that passive channel charting outperforms a baseline based on classical triangulation in terms of localization accuracy.
We discuss our results and highlight some unsolved issues related to the proposed concept.
\end{abstract}

\section{Introduction}
Channel charting is a dimensionality reduction technique that aims to represent the state of a wireless channel in a low-dimensional representation known as the channel chart \cite{studer_cc}.
In a static environment with a mobile \ac{UE} and \acp{BS} at fixed locations, the instantaneous state of the physical wireless channel is entirely determined by the location and orientation of the \ac{UE}.
For sufficiently many \ac{BS} antennas, each \ac{UE} location and orientation $\mathbf x \in \mathbb R^{D'}$ results in a unique high-dimensional \ac{CSI} feature vector $\mathbf f \in \mathbb R^D$ (in arbitrary representation), implying that the mapping from \ac{CSI} to \ac{UE} location and orientation
\[
    \mathcal C_\theta: \mathcal H \to \mathbb R^{D'} ~ \text{with} ~ \mathcal H \subset \mathbb R^D
\]
known as the \ac{FCF}, is injective.
With channel charting, we try to reconstruct $\mathcal C_\theta$, often as a \ac{NN}, by making use of the idea that similarity relationships are preserved between physical space and \ac{CSI} feature space:
Two \ac{CSI} measurements with similar feature vectors $\mathbf f_1, \mathbf f_2 \in \mathbb R^D$ (with respect to some suitable \emph{dissimilarity metric} \cite{stephan2023angle}) likely also belong to two similar \ac{UE} location and orientation vectors $\mathbf x_1, \mathbf x_2 \in \mathbb R^{D'}$.

\begin{figure}
    \centering
    \includegraphics[width=0.75\columnwidth]{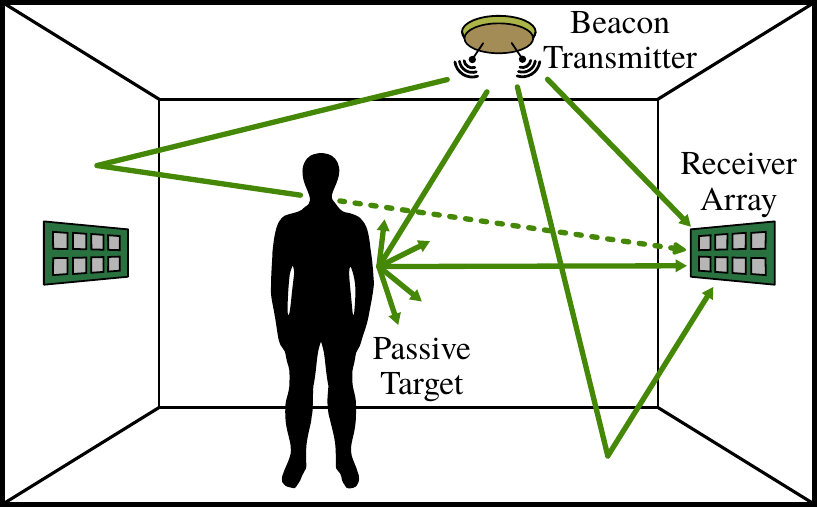}
    \vspace{-0.15cm}
    \caption{Concept of passive channel charting as developed in this work: Passive antenna arrays use signals from non-cooperative beacon transmitter to locate passive target (here: human) in environment, which perturbs clutter channel by scattering / reflecting or absorbing signal components.}
    \label{fig:concept}
    \vspace{-0.4cm}
\end{figure}

Depending on the technique, the learned low-dimensional representation may be directly interpretable in terms of absolute physical quantities (e.g., \ac{UE} location coordinates in meters in a global coordinate frame), or it may only preserve these properties in a relative sense.
In the context of this work, we are concerned with channel charting for two-dimensional localization in absolute coordinates.
In contrast to conventional techniques like triangulation or multilateration, which assume a \ac{LoS} channel between \ac{UE} and \ac{BS} antennas, channel charting does not make any such model assumptions and is thus also feasible in challenging non-\ac{LoS} scenarios \cite{stahlke2023velocity}.

While indoor localization of passive objects based on \mbox{Wi-Fi} \ac{CSI} is a well-investigated subject \cite{ma2019wifi}, to the best of our knowledge, there is only one previous publication applying concepts similar to channel charting to the passive sensing scenario \cite{wicluster}.
We argue and demonstrate\footnote{Partial source code for this work is made publicly available at \texttt{\href{https://github.com/Jeija/ESPARGOS-Passive-ChannelCharting}{github.com/Jeija/ESPARGOS-Passive-ChannelCharting}}} that the underlying principles of channel charting are also applicable to a scenario with immobile transceivers in a static environment that contains a passive (non-transmitting) mobile entity.
Borrowing from radar literature, we call the mobile entity, which may be a  person or an object, the \emph{target} in our system and we will refer to our proposed technique by the name \emph{\ac{PCC}}.
In this radar-like setup, the moving object disturbs the channel between transmitters and receivers, for example by blocking propagation paths or by creating additional reflections (compare Fig.~\ref{fig:concept}).
The resulting perturbation to the \ac{CSI} feature vector $\mathbf f \in \mathbb R^D$ is solely determined by the state $\mathbf x$ of the mobile target (e.g., location and orientation).
As done in conventional channel charting, it should be possible to identify similar target locations and orientations from \ac{CSI} available at the receivers using a suitable dissimilarity metric.
Based on these dissimilarities, a dimensionality reduction algorithm can reconstruct a low-dimensional map of the target states $\mathbf x$, which we will call (passive) channel chart.
Active and passive channel charting have similar benefits: Most importantly, model assumptions (e.g., \ac{LoS} channel, target model) are not required, though they can optionally be incorporated.

\subsection{Related Work}
The authors of \cite{wicluster} already applied dimensionality reduction to passive target localization with Wi-Fi CSI, without calling their approach ``channel charting''.
We propose a similar method to theirs, but now taking into account results from more recent advances in wireless channel charting, such as the use of Siamese neural networks and fused dissimilarity metrics.
In addition, we also provide a classical triangulation-based baseline.
We discussed the \ac{PCC} concept and coordinated this publication with the authors of \cite{stahlke2025passive}, who apply \ac{PCC} to channel measurements acquired from an ultra-wideband system.
Our \ac{PCC} scenario can be interpreted as an implementation of an \ac{ISAC} system, since we rely on conventional Wi-Fi communication signals for target localization.
It may also be seen as a \emph{multi-static radar} system, or even as a \emph{passive radar} system, since \ac{PCC} is also applicable to non-cooperative transmitters.
There is extensive literature on Wi-Fi based passive target sensing, often employing \acp{NN} for localization \cite{ma2019wifi}.
We want to highlight that \ac{PCC} is different to these supervised approaches that employ \ac{CSI} \emph{fingerprinting} in the sense that we do not require labeled training data thanks to our self-supervised approach.

\subsection{Limitations}
We focus on a scenario with a \emph{single} moving target in an otherwise static environment.
While the approach may later be extended to the multi-target case if the targets are separable in some domain (e.g., angle, range, Doppler), our focus is on presenting algorithmic foundations for \ac{PCC} in a simple setup where we can compare \ac{PCC} to a classical baseline.

%\subsection{Overview}
%We first introduce our Wi-Fi channel sounder and the system model in Sec.~\ref{sec:setup} and introduce the clutter removal steps used for all baselines and for \ac{PCC} in Sec.~\ref{sec:clutter}.
%We introduce a classical baseline using triangulation in Sec.~\ref{sec:classical} and a baseline using an \ac{NN} trained in a supervised manner in Sec.~\ref{sec:fingerprinting}.
%\ac{PCC} is implemented using a Siamese \ac{NN} in Sec.~\ref{sec:channelcharting} and compared against the baselines in Sec.~\ref{sec:evaluation}.

\section{System Setup and Dataset}
\label{sec:setup}

\begin{figure}
    \centering
    \includegraphics[width=\columnwidth]{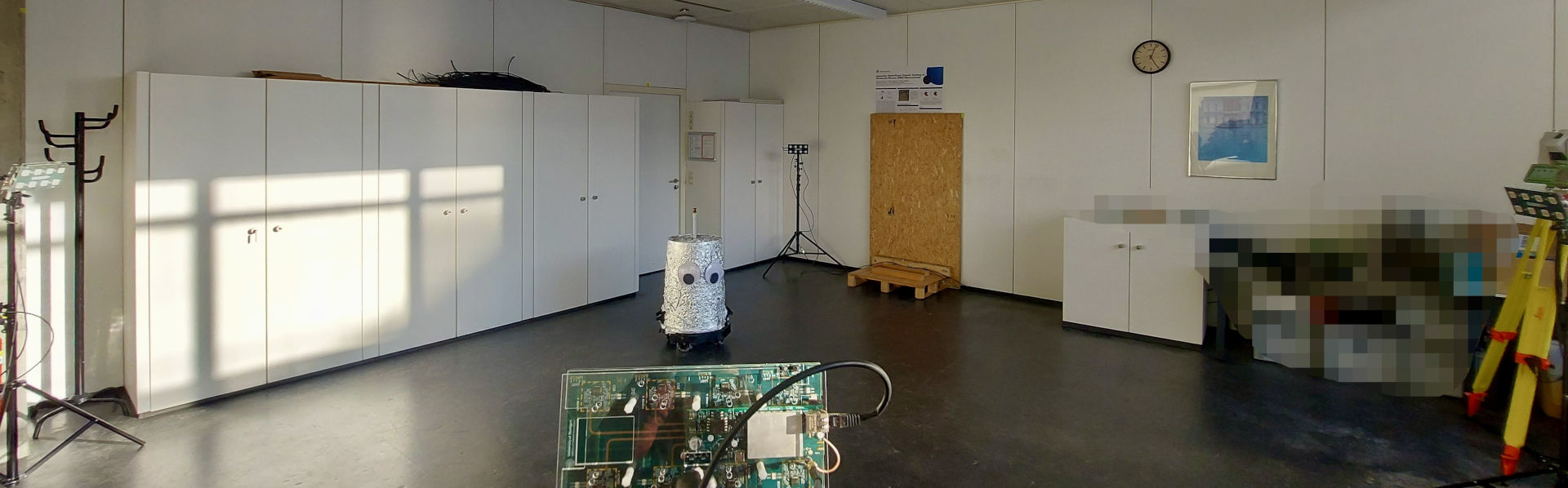}
    \caption{Photo of the environment with the four ESPARGOS arrays (one in foreground, two at the left / right edges and one in the background), the passive target (robot wrapped in aluminium foil) in the middle of the measurement area. Beacon transmitters are mounted to the ceiling and not visible. The dimensions of the measurement area are approximately $4.5\,\mathrm{m} \times 4.5\,\mathrm{m}$.}
    \label{fig:photo}
    \vspace{-0.4cm}
\end{figure}

ESPARGOS \cite{espargos} is a Wi-Fi channel sounder developed at our institute.
It passively acquires \ac{CSI} of Wi-Fi packets through OFDM channel estimation.
The dataset called \mbox{\emph{espargos-0007}}~\cite{dataset-espargos-0007} that is used in this work has been made publicly available.
We use two different types of passive targets: A robot wrapped in aluminium foil to increase the amount of reflected and scattered signal energy, or a human.
The target moves around in the measurement area with fixed upright orientation.
There are $N_\mathrm{TX} = 4$ ceiling-mounted Wi-Fi transmitters and four ESPARGOS arrays made up of $2 \times 4$ antennas each (compare Fig.~\ref{fig:photo}).
While all of the receivers are synchronized in frequency, time, and phase, the transmitters are neither synchronized to each other nor to the receivers and could be interpreted as non-cooperative access points.

We interpret the dataset $\mathcal S$ as a collection of $L$ \emph{datapoints}
\[
    \mathcal S = \{ (\mathbf H^{(l)}, \mathbf x^{(l)}, t^{(l)}, i_\mathrm{TX}^{(l)}) \}_{l = 1, \ldots, L},
\]
where $\mathbf x^{(l)} \in \mathbb R^3$ is the three-dimensional position of the target (in meters, with height coordinate assumed to be known), $t^{(l)} \in \mathbb R$ is a timestamp (in seconds) and $i_\mathrm{TX}^{(l)} \in \mathbb \{1, \ldots, N_\mathrm{TX}\}$ is the index of the transmitter that sent the Wi-Fi packet for which the \ac{CSI} array was acquired.
The CSI array has dimensions $\mathbf H^{(l)} \in \mathbb C^{B \times M_\mathrm{r} \times M_\mathrm{c} \times N_\mathrm{sub}}$, where $B = 4$ is the number of ESPARGOS arrays, $M_\mathrm{r} = 2$ is the number of antenna rows per array, $M_\mathrm{c} = 4$ is the number of antenna columns per array and $N_\mathrm{sub} = 53$ is the number of nonzero subcarriers in the \ac{L-LTF} field of the Wi-Fi preamble used for channel estimation.
The carrier frequency is $f_\mathrm{c} = 2.472\,\mathrm{GHz}$ (Wi-Fi channel 13) and the bandwidth of the signal is $W \approx 16.56\,\mathrm{MHz}$ (true bandwidth of ``$20\,\mathrm{MHz}$'' Wi-Fi channel excluding guard subcarriers).

Furthermore, the position and orientation of all ESPARGOS arrays $b \in \{ 1, \ldots, B \}$ is known and can be described by the array center positions $\mathbf z^{(b)}$ and the array boresight vectors $\mathbf n^{(b)}$.
We want to stress that, in the subsequent sections, both the triangulation baseline and \ac{PCC} use the target position labels $\mathbf x^{(l)}$ only for evaluation.
Only the supervised \ac{NN} baseline (fingerprinting) uses them for training.
We split the dataset into three subsets: A training set with the robot as the target containing $|\mathcal S_\mathrm{rob,train}| = 482882$ datapoints, a test set with the robot containing $|\mathcal S_\mathrm{rob,test}| = 139427$ datapoints and a test set with human target containing $|\mathcal S_\mathrm{hum,test}| = 33011$ datapoints.

\section{Preprocessing: Clutter Map and Clustering}
\label{sec:clutter}
A common challenge in radar systems is dealing with \emph{clutter}, that is, signal components reflected by or scattered off objects other than the target, including direct path components.
The latter is particularly prominent in our scenario with mostly omnidirectional antennas and clear \ac{LoS} paths between transmitters and receivers.
The idea of \emph{clutter removal} is to identify the clutter components of a signal and to subtract them from the target reflection.
While conceptually simple, the lack of phase, time and frequency synchronization between transmitters and receivers makes clutter removal challenging.
Recently, an algorithm called \ac{CRAP} \cite{henninger2023crap} has been proposed for clutter removal under such circumstances.
We briefly outline our implementation of \ac{CRAP}, which we apply separately for each of the four transmitters, but refer the reader to \cite{henninger2023crap, henninger2024augmenting} for details:
\ac{CRAP} works with vectorized \ac{CSI} arrays $\mathrm{vec}\,\mathbf H^{(l)} = \mathbf h^{(l)} \in \mathbb C^{Q}$, where $Q = B \cdot M_\mathrm{r} \cdot M_\mathrm{c} \cdot N_\mathrm{sub}$.
The clutter subspace $\hat {\mathbf C} \in \mathbb C^{Q \times K}$ is determined from the eigenvectors of the autocovariance matrix $\mathbf R = \sum_{l} \mathbf h^{(l)} \left(\mathbf h^{(l)} \right)^\mathrm{H}$
belonging to the $K$ largest eigenvalues, where $K$ is called clutter order.
Note that while \ac{CRAP} assumes an empty room (without target) for clutter acquisition, we apply it to measurements with the target present, but moving around. %, hoping that the influence of the target is thus negligible.
To remove the clutter from a \ac{CSI} measurement $\mathrm{vec}\,\mathbf H^{(l)} = \mathbf h^{(l)}$, we subtract the projected clutter component from the measurement:
\[
    \mathbf h_\mathrm{tgt}^{(l)} = \mathbf h^{(l)} - \hat {\mathbf C} \left( \hat {\mathbf C}^\mathrm{H} \mathbf h^{(l)} \right)
\]
Finally, $\mathbf h_\mathrm{tgt}^{(l)}$ can be re-shaped back into an array $\mathbf H_\mathrm{tgt}^{(l)} \in \mathbb C^{B \times M_\mathrm{r} \times M_\mathrm{c} \times N_\mathrm{sub}}$, which is a clutter-rejected version of the \ac{CSI} that describes the change to the channel due to the target, but is subject to thermal noise, phase noise and other errors.
In our scenario, clutter acquisition and removal has to be performed separately for each transmitter $i_\mathrm{TX}^{(l)}$.
All subsequent chapters use \ac{CSI} with clutter already removed using \ac{CRAP}.

In addition to clutter removal, we perform a clustering step: Instead of working with individual datapoints, we combine all datapoints measured within $\Delta t = 1\,\mathrm{s}$ intervals into a cluster.
Every cluster contains \ac{CSI} from all four transmitters.
We define the set $\mathcal A^{(c)}$ that contains the datapoint indices $l$ of the $c$-th cluster, and denote the mean timestamp for this cluster by $\bar t^{(c)} = \frac{1}{|\mathcal A^{(c)}|} \sum_{l \in \mathcal A^{(c)}} t^{(l)}$ and the mean position label of the cluster by $\mathbf {\bar x}^{(c)} = \frac{1}{|\mathcal A^{(c)}|} \sum_{l \in \mathcal A^{(c)}} \mathbf x^{(l)}$.
A top view map of the position labels $\mathbf {\bar x}^{(c)}$ for dataset $\mathcal S_\mathrm{rob,test}$ is shown in Fig.~\ref{fig:groundtruth}.

\section{Baseline: Classical Triangulation}
\label{sec:classical}

%Since multilateriation using time differences of arrival would be inaccurate due to the small bandwidth, we propose triangulation as a classical, model-based baseline.
We use triangulation as a model-based baseline.
In a first step, we use clutter-rejected datapoints from all transmitters to estimate a cluster-wise azimuth array covariance matrix
\[
    \mathbf R^{(c,b)} = \sum_{l \in \mathcal A^{(c)}} \sum_{m_\mathrm{r} = 1}^{M_\mathrm{r}} \sum_{n = 1}^{N_\mathrm{sub}} \left(\mathbf H_{\mathrm{tgt},b,m_\mathrm{r},:,n}^{(l)}\right) \left(\mathbf H_{\mathrm{tgt},b,m_\mathrm{r},:,n}^{(l)}\right)^\mathrm{H},
\]
where a colon (:) index indicates taking all elements along the corresponding axis of the array.
We determine the azimuth \ac{AoA} $\hat \alpha^{(c,b)}$ for cluster $c$ and array $b$ from $\mathbf R^{(c,b)}$ using the root-MUSIC algorithm assuming a single source, though simpler approaches are also possible.

%\subsection{Triangulation}
In a second step, as in \cite{asilomar2023}, we derive a likelihood function under the assumption of \emph{von Mises}-distributed angle errors.
We denote by $\angle_\mathrm{az}(\mathbf x - \mathbf z^{(b)}, \mathbf n^{(b)})$ the azimuth angle between $\mathbf x - \mathbf z^{(b)}$ (the presumed target position relative to antenna array $b$) and $\mathbf n^{(b)}$ (the normal vector of antenna array $b$).
With $I_0$ denoting the modified Bessel function of the first kind of order 0, the \ac{AoA} likelihood function is
\begin{equation}
    \mathcal L_\mathrm{tri}^{(c)}(\mathbf x) = \prod_{b = 1}^B \frac{\exp \left( \kappa^{(c, b)} \cos \left( \angle_\mathrm{az}(\mathbf x - \mathbf z^{(b)}, \mathbf n^{(b)}) - \hat \alpha^{(c, b)} \right) \right)}{2 \pi I_0(\kappa^{(c, b)})},
    \label{eq:aoalikelihood}
\end{equation}
where $\kappa^{(c, b)}$ is a concentration parameter, which is heuristically derived from the magnitude of the root found with root-MUSIC.
The target position estimate $\mathbf { \hat x }^{(c)}$ is then obtained by numerical optimization of $\mathbf { \hat x }^{(c)} = \argmax_{\mathbf x} \mathcal L^{(c)}_\mathrm{tri}(\mathbf x)$.

We acknowledge that our model-based baseline could potentially be improved even further by exploiting time and phase of arrival information or by exploiting the Doppler effect \cite{li2017indotrack}.

\section{Baseline: Fingerprinting With Neural Network}
\label{sec:fingerprinting}

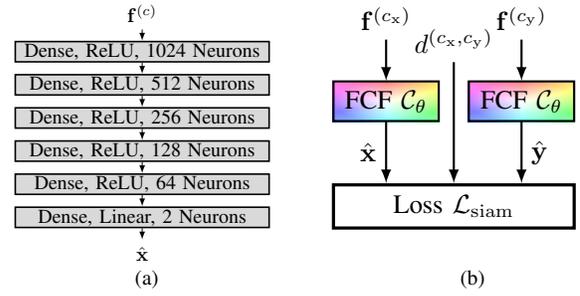
\begin{figure}
    \centering
    \begin{subfigure}[b]{0.5\columnwidth}
        \centering
        \scalebox{0.75}{
            \begin{tikzpicture}
                \node (input) [anchor = south] {$\mathbf f^{(c)}$};
            	\node (l1) [minimum width = 4.5cm, draw, thick, inner sep = 1pt, below = 0.2cm of input, fill = black!15!white] {Dense, ReLU, 1024 Neurons};
            	\node (l2) [minimum width = 4.5cm, draw, thick, inner sep = 1pt, below = 0.2cm of l1, fill = black!15!white] {Dense, ReLU, 512 Neurons};
            	\node (l3) [minimum width = 4.5cm, draw, thick, inner sep = 1pt, below = 0.2cm of l2, fill = black!15!white] {Dense, ReLU, 256 Neurons};
            	\node (l4) [minimum width = 4.5cm, draw, thick, inner sep = 1pt, below = 0.2cm of l3, fill = black!15!white] {Dense, ReLU, 128 Neurons};
            	\node (l5) [minimum width = 4.5cm, draw, thick, inner sep = 1pt, below = 0.2cm of l4, fill = black!15!white] {Dense, ReLU, 64 Neurons};
            	\node (l6) [minimum width = 4.5cm, draw, thick, inner sep = 1pt, below = 0.2cm of l5, fill = black!15!white] {Dense, Linear, 2 Neurons};
    
            	\node (output) [anchor = north] at ($(l6.south) + (0, -0.2)$) {$\hat {\mathbf x}$};
    
                \draw [-latex] (input) -- (l1);
                \draw [-latex] (l1) -- (l2);
                \draw [-latex] (l2) -- (l3);
                \draw [-latex] (l3) -- (l4);
                \draw [-latex] (l4) -- (l5);
            	\draw [-latex] (l5) -- (l6);
            	\draw [-latex] (l6) -- (output);
            \end{tikzpicture}
        }
        \vspace{-0.2cm}
        \caption{}
        \label{fig:nn-structure}
    \end{subfigure}
    \begin{subfigure}[b]{0.45\columnwidth}
            \scalebox{1.0}{
            \begin{tikzpicture}
                \node (in_1) at (2.1,0) {$\mathbf{f}^{(c_\mathrm{x})}$};
                \node (in_2) at (3.9,0) {$\mathbf{f}^{(c_\mathrm{y})}$};
                
                \node (in_3) at (3,-0.3) {$d^{(c_\mathrm{x},c_\mathrm{y})}$};
                
                \node [minimum width = 1.4cm, minimum height = 0.4cm] (fcf1space) at (2.1,-1.1) {FCF $\mathcal{C}_\Theta$};
                \node [minimum width = 1.4cm, minimum height = 0.4cm] (fcf2space) at (3.9,-1.1) {FCF $\mathcal{C}_\Theta$};

                \shade [shading=color wheel white center] (fcf1space.north west) rectangle (fcf1space.south east);
        		\fill [white, opacity = 0.5] (fcf1space.north west) rectangle (fcf1space.south east);

        		\shade [shading=color wheel white center] (fcf2space.north west) rectangle (fcf2space.south east);
        		\fill [white, opacity = 0.5] (fcf2space.north west) rectangle (fcf2space.south east);

                \node [draw, thick, minimum width = 1.4cm, minimum height = 0.4cm] (fcf1) at (2.1,-1.1) {FCF $\mathcal C_\theta$};
		        \node [draw, thick, minimum width = 1.4cm, minimum height = 0.4cm] (fcf2) at (3.9,-1.1) {FCF $\mathcal C_\theta$};

                \node [draw, very thick, black, minimum width = 3.2cm, inner sep = 4pt] (contrastive_loss) at (3,-2.5) {Loss $\mathcal{L}_\mathrm{siam}$};
                
                \draw [-latex, thick]  (in_1.south) -- (fcf1space.north)
                node[midway,anchor=east]{};
                \draw [-latex, thick]  (in_2.south) -- (fcf2space.north)
                node[midway,anchor=east]{};
                
                \draw [-latex, thick]  (in_3.south) -- (contrastive_loss)
                node[midway,anchor=east]{};
                
                \draw [-latex, thick]  (fcf1space.south) -- (fcf1space.south|-contrastive_loss.north)
                node[midway,anchor=east]{$\hat{\mathbf{x}}$};
                \draw [-latex, thick]  (fcf2space.south) -- (fcf2space.south|-contrastive_loss.north)
                node[midway,anchor=west]{$\hat{\mathbf{y}}$};
            \end{tikzpicture}
        }
        \vspace{0.3cm}
        \caption{}
        \label{fig:siamese}
    \end{subfigure}
    \vspace{-0.2cm}
    \caption{Neural network structure: (a) Dense \ac{NN} used for fingerprinting or as \ac{FCF} and (b) \acp{FCF} in Siamese configuration for channel charting training.}
    \vspace{-0.4cm}
\end{figure}

\begin{figure*}
    \centering
    \begin{subfigure}{0.19\textwidth}
        \centering
		\begin{tikzpicture}
			\begin{axis}[
				width=0.73\textwidth,
				height=0.73\textwidth,
				scale only axis,
				xmin=-1.875,
				xmax=3.422,
				ymin=0.403,
				ymax=6.125,
				xlabel = {Coordinate $x_1 ~ \text{in m}$},
				ylabel = {Coordinate $x_2 ~ \text{in m}$},
				ylabel shift = -6 pt,
				xlabel shift = -4 pt
			]
				\addplot[thick,blue] graphics[xmin=-1.875,ymin=0.403,xmax=3.422,ymax=6.125] {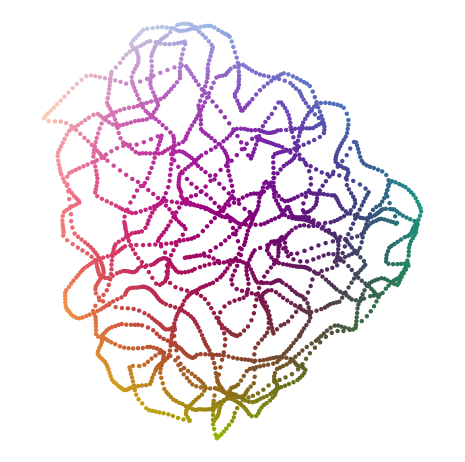};
			\end{axis}
		\end{tikzpicture}
		\vspace{-0.6cm}
		\caption{Position Labels}
		\label{fig:groundtruth}
    \end{subfigure}
    \begin{subfigure}{0.19\textwidth}
        \centering
    	\begin{tikzpicture}
			\begin{axis}[
				width=0.73\textwidth,
				height=0.73\textwidth,
				scale only axis,
				xmin=-1.875,
				xmax=3.422,
				ymin=0.403,
				ymax=6.125,
				xlabel = {Coordinate $\hat x_1 ~ \text{in m}$},
				ylabel = {Coordinate $\hat x_2 ~ \text{in m}$},
				ylabel shift = -6 pt,
				xlabel shift = -4 pt
			]
				\addplot[thick,blue] graphics[xmin=-1.875,ymin=0.403,xmax=3.422,ymax=6.125] {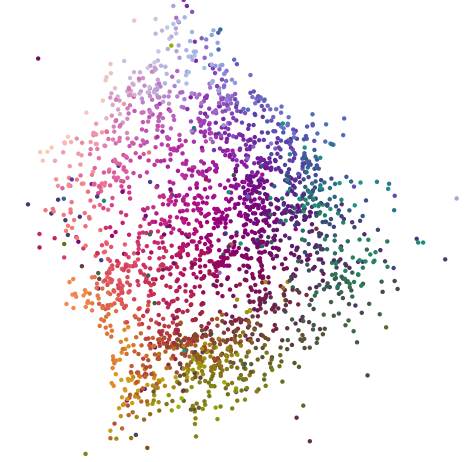};
			\end{axis}
		\end{tikzpicture}
		\vspace{-0.6cm}
		\caption{Triangulation}
		\label{fig:triangulation}
    \end{subfigure}
    \begin{subfigure}{0.19\textwidth}
        \centering
		\begin{tikzpicture}
			\begin{axis}[
				width=0.73\textwidth,
				height=0.73\textwidth,
				scale only axis,
				xmin=-1.875,
				xmax=3.422,
				ymin=0.403,
				ymax=6.125,
				xlabel = {Coordinate $\hat x_1 ~ \text{in m}$},
				ylabel = {Coordinate $\hat x_2 ~ \text{in m}$},
				ylabel shift = -6 pt,
				xlabel shift = -4 pt
			]
				\addplot[thick,blue] graphics[xmin=-1.875,ymin=0.403,xmax=3.422,ymax=6.125] {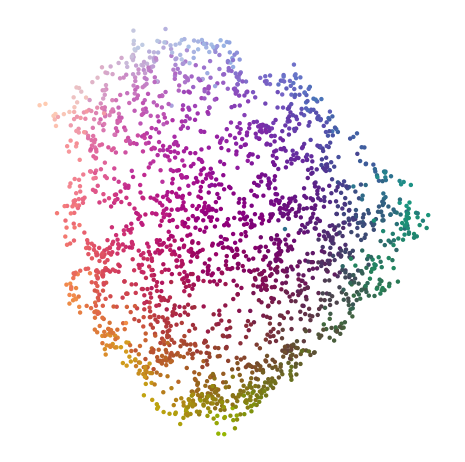};
			\end{axis}
		\end{tikzpicture}
		\vspace{-0.6cm}
		\caption{Fingerprinting}
		\label{fig:fingerprinting}
    \end{subfigure}
    \begin{subfigure}{0.19\textwidth}
        \centering
		\begin{tikzpicture}
			\begin{axis}[
				width=0.73\textwidth,
				height=0.73\textwidth,
				scale only axis,
				xmin=-2.617,
				xmax=2.768,
				ymin=-2.612,
				ymax=2.623,
				xlabel = {Coordinate $\hat x_1$},
				ylabel = {Coordinate $\hat x_2$},
				ylabel shift = -6 pt,
				xlabel shift = -4 pt
			]
				\addplot[thick,blue] graphics[xmin=-2.617,ymin=-2.612,xmax=2.768,ymax=2.623] {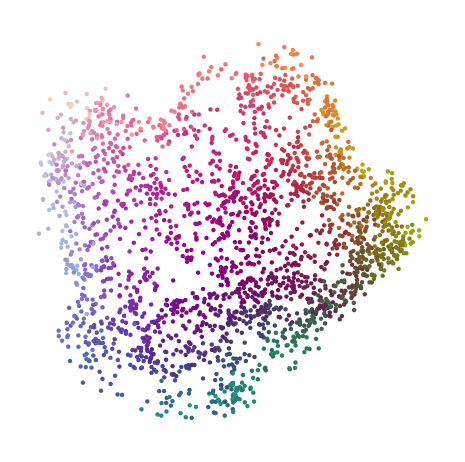};
			\end{axis}
		\end{tikzpicture}
		\vspace{-0.6cm}
		\caption{PCC $\mathcal C_\theta$}
		\label{fig:pcc}
    \end{subfigure}
    \hspace{0.1cm}
    \begin{subfigure}{0.19\textwidth}
        \centering
		\begin{tikzpicture}
			\begin{axis}[
				width=0.73\textwidth,
				height=0.73\textwidth,
				scale only axis,
				xmin=-1.875,
				xmax=3.422,
				ymin=0.403,
				ymax=6.125,
				xlabel = {Coordinate $\hat x_1 ~ \text{in m}$},
				ylabel = {Coordinate $\hat x_2 ~ \text{in m}$},
				ylabel shift = -6 pt,
				xlabel shift = -4 pt
			]
				\addplot[thick,blue] graphics[xmin=-1.875,ymin=0.403,xmax=3.422,ymax=6.125] {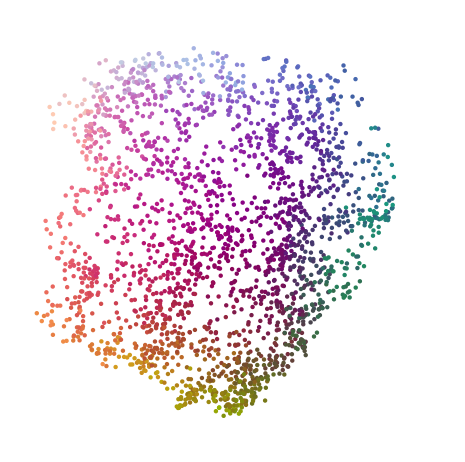};
            \end{axis}
		\end{tikzpicture}
		\vspace{-0.6cm}
		\caption{Augmented \ac{PCC} $\mathcal C_{\theta,\mathrm{aug}}$}
		\label{fig:augmented}
    \end{subfigure}
    \vspace{-0.3cm}
    \caption{Top view map of colorized position labels in $\mathcal S_\mathrm{rob,test}$ after clustering shown in (a). Datapoint colors are preserved for (b), (c), (d) and (e), which show (b) position estimates produced by the classical triangulation baseline (c) position estimates from the supervised \ac{NN}, (d) the passive channel chart before the coordinate transform and (e) the augmented passive channel chart, with classical \ac{AoA} estimates considered during training (all evaluated on $\mathcal S_\mathrm{rob,test}$).}
    \label{fig:channelcharts}
    \vspace{-0.5cm}
\end{figure*}

One of the simplest machine learning-based methods for indoor localization of transmitters or passive targets is \ac{CSI} fingerprinting \cite{wang2016csi, ma2019wifi, zhou2018device}.
It assumes that a training set containing \ac{CSI} with associated position labels exists that a machine learning model, such as an \ac{NN}, can be trained on.
While \ac{CSI} fingerprinting usually provides good localization performance, it requires a labeled training dataset, which is often unavailable, and generalizes poorly to previously unseen environments and target types.
We train the \ac{NN} regression model shown in Fig.~\ref{fig:nn-structure}, which directly predicts target positions $\mathbf x$ in a supervised manner with \ac{MSE} loss.
Instead of providing raw \ac{CSI} matrices $\mathbf H_\mathrm{tgt}^{(l)}$ to the \ac{NN}, we use suitable feature engineering to design features that are more easily processed.
We compute the features as follows:
First, the frequency-domain clutter-rejected \ac{CSI} arrays $\mathbf H_\mathrm{tgt}^{(l)}$ are transformed to time domain by taking the \ac{FFT} over the subcarrier axis.
To reduce the number of input features, only the $N_\mathrm{tap} = 12$ time taps (taps 22 to 34) containing meaningful signal components are extracted, yielding time-domain \ac{CSI} arrays $\mathbf {H'}_\mathrm{tgt}^{(l)} \in \mathbb C^{B \times M_\mathrm{r} \times M_\mathrm{c} \times N_\mathrm{tap}}$.
Next, we compute feature arrays $\mathbf F^{(c)}_{i_\mathrm{TX}, b, t} \in \mathbb C^{M_\mathrm{r} \cdot M_\mathrm{c}\times M_\mathrm{r} \cdot M_\mathrm{c}}$ for every cluster $c$, transmitter $i_\mathrm{TX}$, array $b$ and time tap $t$ as
\[
    \mathbf F^{(c)}_{i_\mathrm{TX}, b, t} = \sum_{\substack{l \in \mathcal A^{(c)}\\ i_\mathrm{TX}^{(l)} = i_\mathrm{TX}}} \left( \mathrm{vec} \, \mathbf {H'}_{\mathrm{tgt},b,:,:,t}^{(l)} \right) \left( \mathrm{vec} \, \mathbf {H'}_{\mathrm{tgt},b,:,:,t}^{(l)} \right)^\mathrm{H},
\]
where the condition $i_\mathrm{TX}^{(l)} = i_\mathrm{TX}$ means that the sum is only computed over those datapoints in the cluster containing \ac{CSI} for transmitter $i_\mathrm{TX}$.
Finally, we vectorize feature arrays $\mathbf F^{(c)}_{i_\mathrm{TX}, b, t}$ for a particular cluster $c$ and provide their real and imaginary parts to the \ac{NN} in a feature vector $\mathbf f^{(c)} \in \mathbb R^{2 \cdot N_\mathrm{TX} \cdot B \cdot N_\mathrm{tap} \cdot M_\mathrm{r}^2 \cdot M_\mathrm{c}^2}$.

\section{Passive Channel Charting}
\label{sec:channelcharting}

In previous work on channel charting, various dimensionality reduction techniques have been applied to \ac{CSI} datasets.
In the following, we focus on dissimilarity metric-based channel charting with a Siamese \ac{NN}, which has shown good performance for localization tasks with reasonable computational complexity \cite{stephan2023angle, fraunhofer_cc}.
This means that we first need to compute pseudo-distances between datapoints called dissimilarities.

\subsection{Dissimilarity Metric Construction}
We extend the definition of the cosine similarity-based dissimilarity introduced in \cite{le2021efficient} to the passive target case with multiple arrays as follows:
First, we combine all clutter-rejected \ac{CSI} measurements $\mathbf H_\mathrm{tgt}^{(l)}$ belonging to the same cluster into a single \ac{CSI} matrix $\mathbf {\bar H}_\mathrm{tgt}^{(l)} \in \mathbb C^{B \times M_\mathrm{r} \times M_\mathrm{c}}$ using a subspace-based interpolation method that superimposes the contributions of all subcarriers.
We then straightforwardly apply the cosine similarity-based dissimilarity metric as
\[
    d_{\mathrm{CS}}^{(i,j)} = B - \sum_{b = 1}^{B} \sum_{m_\mathrm{r} = 1}^{M_\mathrm{r}} \sum_{m_\mathrm{c} = 1}^{M_\mathrm{c}} \ \frac{\left| \left(\mathbf {\bar H}_{\mathrm{tgt},b,m_\mathrm{r},m_\mathrm{c}}^{(i)}\right)^* \mathbf {\bar H}_{\mathrm{tgt},b,m_\mathrm{r},m_\mathrm{c}}^{(j)} \right|^2}{\left\lVert \mathbf {\bar H}_{\mathrm{tgt},b}^{(i)} \right\rVert_\mathrm{F}^2 \left\lVert \mathbf {\bar H}_{\mathrm{tgt},b}^{(j)} \right\rVert_\mathrm{F}^2},
\]
where $\lVert \cdot \rVert_\mathrm{F}$ denotes the Frobenius norm.
In contrast to \cite{le2021efficient}, we need to compute sums over both row index $m_\mathrm{r}$ and column index $m_\mathrm{c}$, and also over the array index $b$.
The resulting dissimilarity value $d_\mathrm{CS}^{(i,j)}$ can be interpreted as a pseudo-distance between datapoint clusters $i$ and $j$.
As in \cite{stephan2023angle}, we use the time difference $\left|\bar t^{(i)} - \bar t^{(i)}\right|$ between clusters $i$ and $j$ to create a \emph{fused} dissimilarity $d_{\mathrm{CS-fuse}}^{(i,j)}$ and apply the concept of geodesic dissimilarities highlighted in \cite{fraunhofer_cc} to obtain the geodesic fused dissimilarities $d_{\mathrm{CS-fuse,geo}}^{(i,j)}$ used for training.

\subsection{Neural Network Training}
The \ac{FCF} $\mathcal C_\theta$ is implemented as an \ac{NN} with the same structure as in Sec.~\ref{sec:fingerprinting} (compare Fig.~\ref{fig:nn-structure}).
As before, the \ac{NN} predicts target positions $\mathbf x$ from cluster-wise feature vectors $\mathbf f^{(c)}$, which are calculated as in Sec.~\ref{sec:fingerprinting}.
In contrast to supervised training, however, channel charting trains the \ac{NN} in a self-supervised manner using the Siamese \ac{NN} configuration shown in Fig.~\ref{fig:siamese} and a specially designed Siamese loss function
\begin{equation}
    \mathcal{L}_\mathrm{siam}^{(c_\mathrm{x}, c_\mathrm{y})}(\mathbf {\hat x}, \mathbf {\hat y}) = \frac{\left(d_\mathrm{CS-fuse,geo}^{(c_\mathrm{x}, c_\mathrm{y})} - \Vert\mathbf {\hat y} - \mathbf {\hat x}\Vert_2\right)^2}{d_\mathrm{CS-fuse,geo}^{(c_\mathrm{x}, c_\mathrm{y})} + \beta}.
    \label{eq:siameseloss}
\end{equation}

In Eq.~(\ref{eq:siameseloss}), $\mathbf {\hat x} = \mathcal C_\theta(\mathbf f^{(c_\mathrm{x})})$ and $\mathbf {\hat y} = \mathcal C_\theta(\mathbf f^{(c_\mathrm{y})})$ are the latent space (channel chart) position predictions for the clusters with indices $c_\mathrm{x}$ and $c_\mathrm{y}$.
The parameter $\beta$ is a hyperparameter that can tune Eq.~(\ref{eq:siameseloss}) to weight either the absolute squared error (for large $\beta$) or the normalized squared error (for small $\beta$) higher.
$\mathcal{L}_\mathrm{siam}^{(c_\mathrm{x}, c_\mathrm{y})}$ compares the Euclidean distance of predicted channel chart locations $\lVert \hat { \mathbf y } - \hat { \mathbf x } \rVert_2$ to the computed dissimilarity $d_\mathrm{G-fuse}^{(c_\mathrm{x}, c_\mathrm{y})}$.
Intuitively, the loss function is minimized if a low-dimensional representation (channel chart) is found such that the Euclidean distances in the chart match the computed dissimilarities, which is a common objective in dimensionality reduction.

While the \ac{FCF} ideally preserves relative positions, the channel chart's coordinate frame usually does not to match the global coordinate frame.
As in \cite{fraunhofer_cc}, we evaluate the channel chart after a transform $\mathcal T$ to the physical coordinate frame to compute meaningful localization performance metrics.
It is only for this final evaluation step that we use the position labels $\mathbf x^{(c)}$ to determine the affine transform $\mathcal T_\mathrm{opt}$ that is optimal with respect to the mean squared error between position label $\mathbf x^{(c)}$ and transformed prediction $\mathcal T_\mathrm{opt}(\hat {\mathbf x}^{(c)}) = \mathcal T_\mathrm{opt} \circ \mathcal C_\theta\left(\mathbf f^{(c)}\right)$.

\subsection{Augmented Channel Charting}
Instead of relying on position labels to find $\mathcal T_\mathrm{opt}$, we can also incorporate the classical triangulation approach into the \ac{NN}'s loss function as proposed in \cite{asilomar2023}.
First, we use the position predictions obtained via classical triangulation as in Sec.~\ref{sec:classical} to scale the dissimilarities $d_\mathrm{CS-fuse,geo}^{(i,j)}$ to the same unit as the global coordinate frame, e.g., meters.
Then, we combine Eq.~\ref{eq:aoalikelihood} and Eq.~\ref{eq:siameseloss} into a composite loss function

\begin{equation}
    \begin{split}
        \mathcal{L}_\mathrm{comb}^{(c_\mathrm{x}, c_\mathrm{y})}(\hat {\mathbf x}, \hat {\mathbf y}) = (1 - \lambda) & \mathcal L_\mathrm{siam}^{(c_\mathrm{x}, c_\mathrm{y})}(\hat {\mathbf x}, \hat {\mathbf y}) \\- \lambda &\left(\mathcal L^{(c_\mathrm{y})}_\mathrm{tri}(\hat {\mathbf y}) + \mathcal L^{(c_\mathrm{x})}_\mathrm{tri}(\hat {\mathbf x})\right),
    \end{split}
    \label{eq:combinedloss}
\end{equation}
where $\lambda$ is a hyperparameter that controls the weight of the classical triangulation loss relative to the Siamese loss.
A \ac{FCF} $\mathcal C_{\theta, \mathrm{aug}}$ trained with $\mathcal L_\mathrm{comb}$ directly predicts target positions in the absolute global coordinate frame.

\section{Evaluation and Discussion of Results}
\label{sec:evaluation}

\begin{figure}
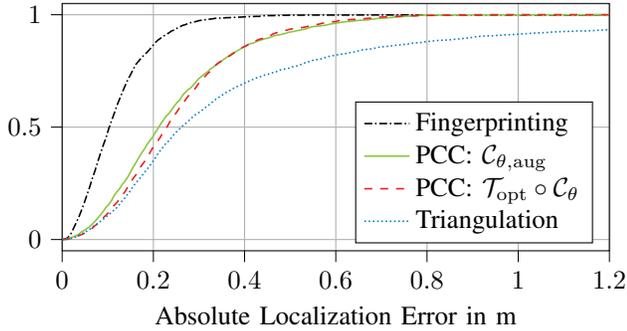

    \centering
    \include{fig/cdf}
    \vspace{-1.1cm}
    \caption{Empirical cumulative distribution functions of absolute localization errors for both baselines and \ac{PCC} (augmented and un-augmented after coordinate transform $\mathcal T_\mathrm{opt}$)}
    \label{fig:ecdf}
    \vspace{-0.4cm}
\end{figure}

\begin{table*}
    \centering
    \captionof{table}{Comparison of typical performance using common localization performance metrics and dimensionality reduction metrics}
    \vspace{-0.1cm}
    \setlength{\tabcolsep}{4pt}
    \renewcommand{\arraystretch}{0.9}
    \begin{tabular}{r c c | c c c c c c | c c}
        & \textbf{Training Set} & \textbf{Test Set} & \textbf{MAE $\downarrow$} & \textbf{DRMS $\downarrow$} & \textbf{CEP $\downarrow$} & \textbf{R95 $\downarrow$} & \textbf{KS $\downarrow$} & \textbf{CT/TW $\uparrow$} & \textbf{Fig.} & \textbf{Graph} \\ \hline
        Baseline: Triangulation & not needed & $\mathcal S_\mathrm{rob,test}$ & $0.434\,\mathrm m$ & $0.694\,\mathrm m$ & $0.261\,\mathrm m$ & $1.368\,\mathrm m$ & $0.292$ & $0.926/0.920$ & \ref{fig:triangulation} & Fig.~\ref{fig:ecdf}: \tikz{\draw[semithick, densely dotted, mittelblau] (0, 0) -- (0.5,0); \node [inner sep = 0.5ex] at (0.25, 0) {};} \\
        Baseline: Supervised \ac{NN} & $\mathcal S_\mathrm{rob,train}$ & $\mathcal S_\mathrm{rob,test}$ & $0.123\,\mathrm m$ & $0.149\,\mathrm m$ & $0.104\,\mathrm m$ & $0.278\,\mathrm m$ & $0.069$ & $0.996/0.996$ & \ref{fig:fingerprinting}  & Fig.~\ref{fig:ecdf}: \tikz{\draw[semithick, densely dashdotted, black] (0, 0) -- (0.5,0); \node [inner sep = 0.5ex] at (0.25, 0) {};} \\
        PCC: $\mathcal T_\mathrm{opt} \circ \mathcal C_\theta$ & $\mathcal S_\mathrm{rob,train}$ & $\mathcal S_\mathrm{rob,test}$ & $0.257\,\mathrm m$ & $0.298\,\mathrm m$ & $0.231\,\mathrm m$ & $0.556\,\mathrm m$ & $0.146$ & $0.986/0.988$ & \ref{fig:pcc} & Fig.~\ref{fig:ecdf}: \tikz{\draw[semithick, dashed, rot] (0, 0) -- (0.5,0); \node [inner sep = 0.5ex] at (0.25, 0) {};} \\
		Augmented PCC: $\mathcal C_{\theta,\mathrm{aug}}$ & $\mathcal S_\mathrm{rob,train}$ & $\mathcal S_\mathrm{rob,test}$ & $0.258\,\mathrm m$ & $0.310\,\mathrm m$ & $0.219\,\mathrm m$ & $0.585\,\mathrm m$ & $0.139$ & $0.985/0.988$ & \ref{fig:augmented} & Fig.~\ref{fig:ecdf}: \tikz{\draw[semithick, apfelgruen] (0, 0) -- (0.5,0); \node [inner sep = 0.5ex] at (0.25, 0) {};} \\ \hline

        Baseline: Triangulation & not needed & $\mathcal S_\mathrm{hum,test}$ & $0.322\,\mathrm m$ & $0.499\,\mathrm m$ & $0.227\,\mathrm m$ & $0.775\,\mathrm m$ & $0.123$ & $0.989/0.988$ & \multicolumn{2}{c}{omitted} \\
        Baseline: Supervised \ac{NN} & $\mathcal S_\mathrm{rob,train}$ & $\mathcal S_\mathrm{hum,test}$ & $0.487\,\mathrm m$ & $0.683\,\mathrm m$ & $0.311\,\mathrm m$ & $1.423\,\mathrm m$ & $0.206$ & $0.961/0.975$ & \multicolumn{2}{c}{omitted} \\
        PCC: $\mathcal T_\mathrm{opt} \circ \mathcal C_\theta$ & $\mathcal S_\mathrm{rob,train}$ & $\mathcal S_\mathrm{hum,test}$ & $0.532\,\mathrm m$ & $0.716\,\mathrm m$ & $0.387\,\mathrm m$ & $1.448\,\mathrm m$ & $0.263$ & $0.935/0.960$ & \multicolumn{2}{c}{omitted} \\
		Augmented PCC: $\mathcal C_{\theta,\mathrm{aug}}$ & $\mathcal S_\mathrm{rob,train}$ & $\mathcal S_\mathrm{hum,test}$ & $0.558\,\mathrm m$ & $0.746\,\mathrm m$ & $0.370\,\mathrm m$ & $1.588\,\mathrm m$ & $0.249$ & $0.951/0.965$ & \multicolumn{2}{c}{omitted} \\
    \end{tabular}
    \label{tab:performance}
    \vspace{0.1cm}

    \textbf{MAE} = Mean Absolute Error, \textbf{DRMS} = Distance Root Mean Square, \textbf{CEP} = Circular Error Probable, \textbf{R95} = 95\textsuperscript{th} percentile radius,\newline
    \textbf{CT} = Continuity, \textbf{TW} = Trustworthiness, \textbf{KS} = Kruskal's Stress, all metrics as defined in \cite{stephan2023angle, asilomar2023}
    \vspace{-0.5cm}
\end{table*}

We train the supervised \ac{NN} baseline and both \ac{PCC} models (un-augmented with loss $\mathcal L_\mathrm{siam}$ and augmented with loss $\mathcal L_\mathrm{comb}$) on training set $\mathcal S_\mathrm{rob, train}$ and evaluate all models as well as the triangulation baseline on both test sets $\mathcal S_\mathrm{rob, test}$ and $\mathcal S_\mathrm{hum, test}$.
We compute the localization and dimensionality reduction performance metrics commonly used in channel charting literature as defined, for example, in \cite{stephan2023angle, asilomar2023}.
The performance metrics are listed in Tab. \ref{tab:performance}.
For test set $\mathcal S_\mathrm{rob, test}$, the corresponding channel charts / position estimates are shown in Fig.~\ref{fig:channelcharts} and the empirical cumulative distribution function of localization errors is shown in Fig.~\ref{fig:ecdf}.

All localization methods deliver convincing results for this scenario.
Unsurprisingly, \ac{CSI} fingerprinting with an \ac{NN} trained in a supervised manner outperforms all other methods, but at the cost of requiring position labels.
\ac{PCC} is more accurate than triangulation, and augmenting \ac{PCC} with classical \ac{AoA} estimates provides position estimates in an absolute global coordinate frame with similar accuracy compared to the un-augmented case combined with the optimal coordinate transform.
A major challenge of all \ac{NN}-based techniques, including \ac{PCC}, is generalization:
Triangulation performs better with a human target compared to the robot as target.
This can likely be explained by the higher \ac{RCS} of the human.
On the other hand, when trained on $\mathcal S_\mathrm{rob, train}$ and evaluated on $\mathcal S_\mathrm{hum, test}$, the performance of \ac{PCC} and the fingerprinting \ac{NN} falls short of classical triangulation, which hints at the problem of overfitting to a specific target type.

\section{Conclusion and Outlook}
We confirmed the feasibility of applying the algorithmic framework of channel charting to the passive target case.
The results indicate that \ac{PCC} can outperform classical localization methods, but at the cost of overfitting to a particular target type, which is an issue that channel charting with an active transmitter does not have.
Future work may look at ways to mitigate this overfitting or find methods to quickly adapt to different targets.
Another open challenge is the extension of \ac{PCC} to multiple targets, which seems feasible in principle.
\ac{PCC} could enable some degree of (partial) non-\ac{LoS} sensing since the dimensionality reduction approach does not make model assumptions.
Our work can only be seen as a confirmation of the attractiveness of \ac{PCC}, with many open questions and interesting opportunities for further research still remaining.

\bibliographystyle{IEEEtran}
\bibliography{IEEEabrv,references}

\end{document}